\documentclass[aps,prl,twocolumn,superscriptaddress,showpacs,10pt]{revtex4-1}

\usepackage{graphicx}
\usepackage[usenames]{color}
\usepackage{ulem}
\definecolor{eli}{cmyk}{0.0, 1.0, 0.8, 0.0} 

\usepackage{graphicx}
\usepackage{amssymb}
\usepackage{epstopdf}
\usepackage{times}
\begin{document}
\author{Young Jun Chang,$^{1,2}$ Aaron Bostwick,$^{1}$ Yong Su Kim,$^{1,3}$ Karsten Horn,$^{2}$ Eli Rotenberg}

\address{ Advanced Light Source, Lawrence Berkeley National Laboratory, Berkeley, California, USA,\\ $^2$ Fritz-Haber-Institut der Max-Planck-Gesellschaft, Berlin, Germany,\\ $^3$ Department of Applied Physics, Hanyang University,Ansan, Gyeonggi-do, 426-791, Korea}







\title{Structure and Correlation Effects in Semiconducting SrTiO$ _{3}$}
\date{\today}

\def\EF{$E_{\mathrm{F}}$}
\def\ED{$E_{\mathrm{D}}$}
\def\TC{$T_{\mathrm{c}}$}

\def\kpar{$k_{\parallel}$}
\def\kperp{$k_{\perp}$}
\def\ky{$k_{y}$}
\def\kx{$k_{x}$}
\def\kKpoint{\kpar$=1.703~$\AA$^{-1}$}
\def\kk{K-K$^\prime$}

\def\dxy{$d_{xy}$}		
\def\dyz{$d_{yz}$}
\def\dzx{$d_{zx}$}

\def\A1{\AA$^{-1}$}
\def\vF{$v_{F}$}
\def\n{$n$}
\def\me{$m^{*}$}
\def\meh{$m^{*}_{h}$}
\def\mel{$m^{*}_{l}$}
\def\m0{$m_{0}$}

\def\ee{\emph{e-e}}
\def\eph{\emph{e-ph}}
\def\epl{\emph{e-pl}}

\def\hv{$h\nu$}	

\def\etal{\textit{et al.}}
\def\vs{\textit{vs.}}
\def\ie{\textit{i.e.}}
\def\eg{\textit{e.g.}}
\def\BZ{Brillouin Zone}

\def\cq{$\chi_0(\mathbf{q})$}

\def\figonecaption{LEED patterns from the TiO$ _{2}$-terminated (001) surfaces of n-type STO at 300$^\circ$K (a) and 20$^\circ$K (b), taken at the beam energy of 235 eV. The arrows mark the newly developed spots at 20$^\circ$K, and the solid lines indicate the principal directions [100] or [010].}
\def\figone{
\begin{figure}[ht] \begin{center}
     {\includegraphics[width=3.375in]{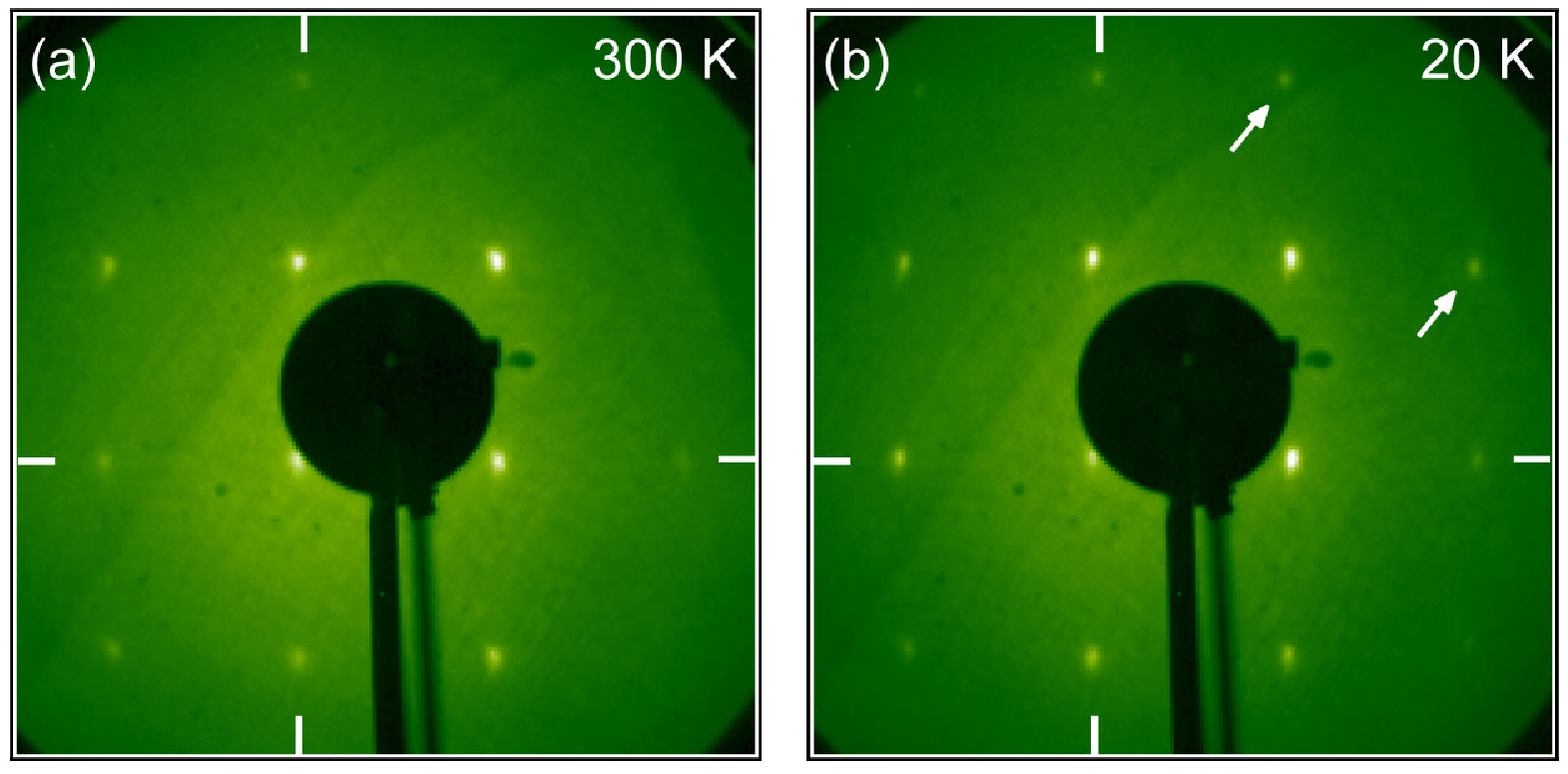}}
     \caption{\figonecaption\label{fig:datasp}}
\end{center}
\end{figure}
 }

\def\figtwocaption{The Fermi surface maps and energy distribution curves of the n-type STO for 150$^\circ$K (a)-(e) and 20$^\circ$K (f)-(j). (a),(f), The Fermi surfaces with schematic pictures as insets describing the shapes of Fermi surface (red solid) in the first \BZ\ boundary (black solid). The Fermi surfaces for (-2$\pi$, -2$\pi$) (b),(g) and (0, -2$\pi$) (d),(i) as indicated with black dashed and dotted boundaries in (a) and (f). The red dashed lines schematically show the Fermi surface contours. The associated band structure cuts for (-2$\pi$, -2$\pi$) (c),(h) and (0, -2$\pi$) (e),(j) along the horizontal black dashed and dotted lines, respectively.
}
\def\figtwo{
\begin{figure}[ht] \begin{center}
     {\includegraphics[width=3.375in]{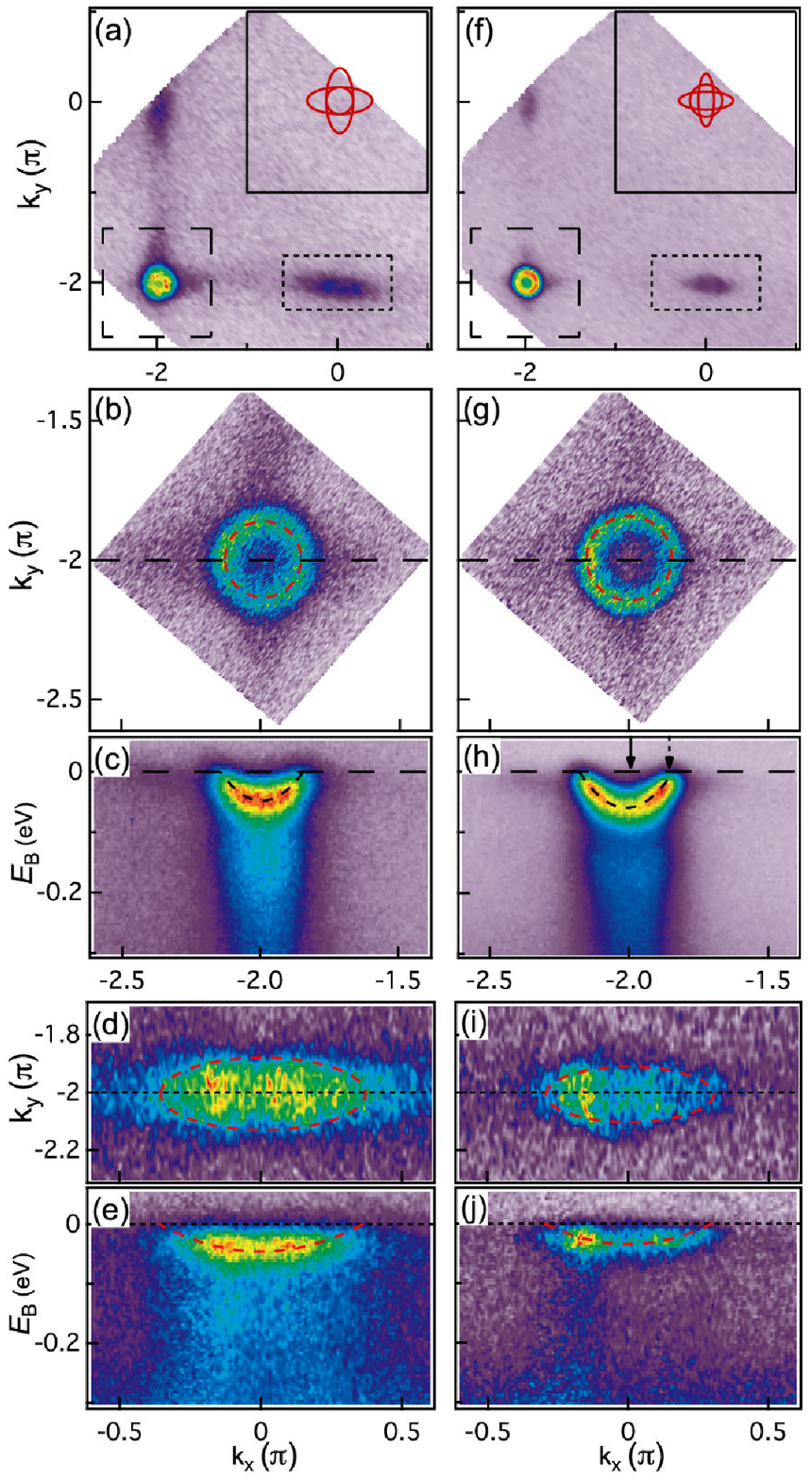}}
     \caption{\figtwocaption\label{fig:datasp}}
\end{center}
\end{figure}
 }

\def\figthreecaption{Energy distribution curves at \kx\ = -2$\pi$ (a) and -1.85$\pi$ (b) of \dxy\ pocket between 20$^\circ$K and 150$^\circ$K, as indicated by solid and dashed arrows in Fig.\ 2(h). The dashed curves at the bottom replicate the energy distribution curves at 150$^\circ$K for better comparison. The arrows mark the hump states at around 170 meV.
}
\def\figthree{
\begin{figure}[ht] \begin{center}
     {\includegraphics[width=3.375in]{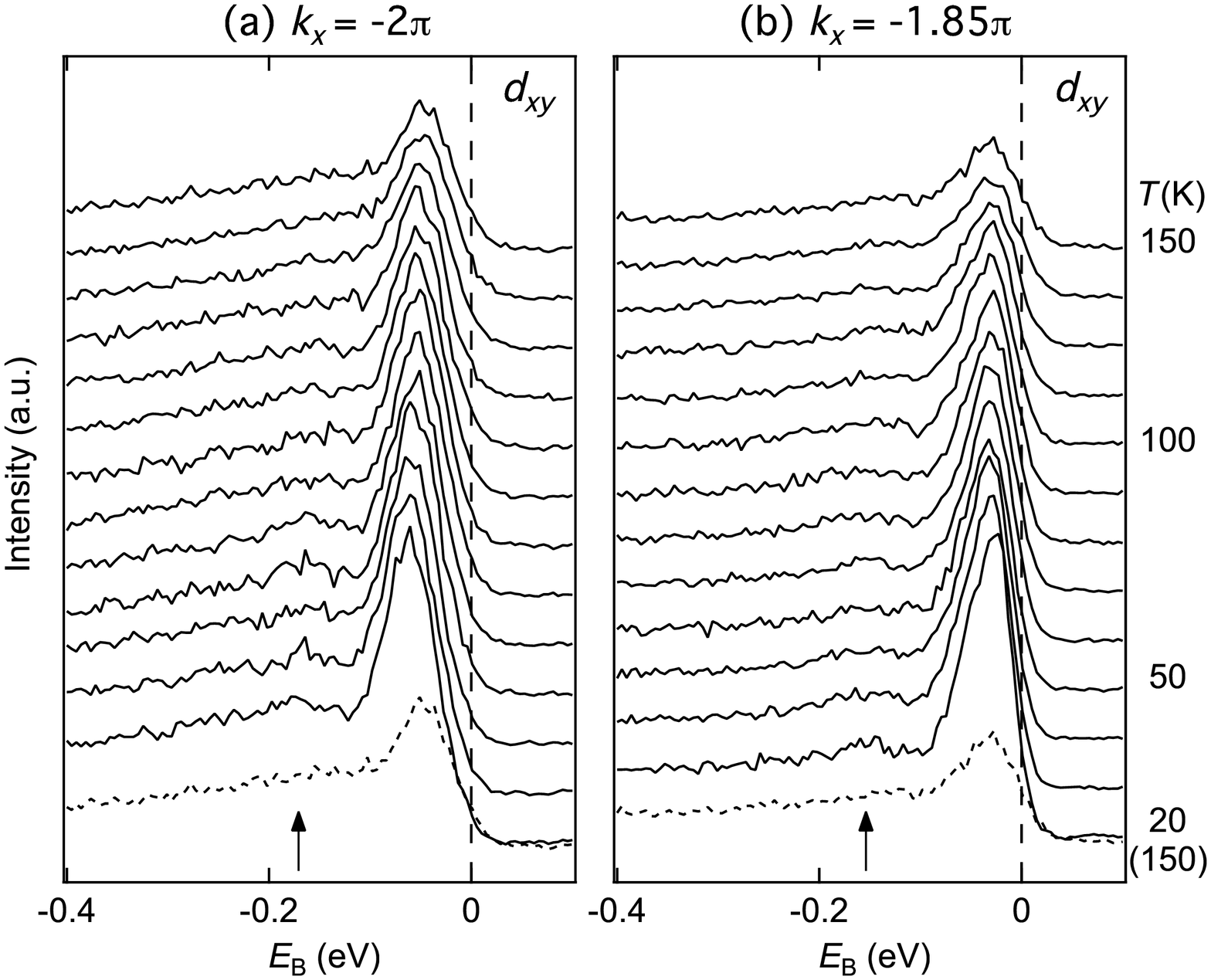}}
     \caption{\figthreecaption\label{fig:datasp}}
\end{center}
\end{figure}
 }

\begin{abstract}

We have investigated the effects of structure change and electron
correlation on SrTiO$ _{3}$ single crystals using angle-resolved
photoemission spectroscopy. We show that the cubic to tetragonal
phase transition at 105$^\circ$K is manifested by a charge transfer from
in-plane ($d_{yz}$ and $d_{zx}$) bands to out-of-plane ($d_{xy}$) band, which is
opposite to the theoretical predictions. Along this second-order
phase transition, we find a smooth evolution of the quasiparticle
strength and effective masses. The in-plane
band exhibits a peak-dip-hump lineshape, indicating a high degree of
correlation on a relatively large (170 meV) energy scale, which is
attributed to the polaron formation.

\end{abstract}
\pacs{71.20.-b, 79.60.-i, 77.80.B-} \maketitle

Among complex oxides, SrTiO$_{3}$ (STO) is one of the most widely
studied examples. Complex oxides present a wide spectrum of
interesting phenomena such as high temperature superconductivity,
colossal magnetoresistance, ferroelectricity, and multiferroicity.
Since many of the interesting oxides have perovskite structure,
STO (a typical cubic perovskite) has been widely used for
integration with other oxides into novel heterostructures. Those
heterostructures show intriguing phenomena such as
superconductivity \cite{caviglia2008, reyren2007, kozuka2009},
high-mobility electron gas \cite{ohtomo2004}, thermoelectricity
\cite{ohta2007, wunderlich2008}, and interface magnetism
\cite{brinkman2007}. Furthermore, STO itself has shown interesting
properties such as superconductivity \cite{ueno2008}, blue-light
emission \cite{kan2005}, photovoltaic effect \cite{zhou2009}, and
water photolysis \cite{konta2004}. Such functionalities would be
improved through the better understanding of the electronic
structure of STO.

STO is a band insulator, which experiences 
a second-order phase transition from cubic to tetragonal structure 
at the critical temperature \TC\ of about 105$^\circ$K \cite{wang1973}.
Below \TC, neighboring octahedral TiO$_{6}$ units start to rotate
oppositely about the $c$ axis eventually reaching an angle of 2$^\circ$
as $T\rightarrow 0$. The increasing tetragonality smoothly changes some
physical properties, such as dielectric constant \cite{muller1979} and
birefringence \cite{courtens1972}. Mattheiss predicted the presence
of three doubly-degenerate Ti 3$d$ bands, \ie\ three orthogonal
ellipsoids centered at the zone center (0,0) \cite{mattheiss1972a}.
During the structural change below \TC, it is predicted that the
degenerate conduction band minimum splits into two; the in-plane
(\dyz\ and \dzx) bands at lower energy and the out-of-plane (\dxy)
band at higher energy \cite{mattheiss1972b}. Both in-plane and out-of-plane bands are 
to be occupied for carrier density \n\ $>$ 10$ ^{19}$ cm$^{-3}$. 

While earlier transport \cite{gregory1979, uwe1985} and angle-resolved photoemission
spectroscopy (ARPES) \cite{aiura2002, takizawa2009, haruyama1996, ishida2008}  measurements  
provided some information, important details such as the Fermi surface size and band offsets are 
in disagreement with theory \cite{mattheiss1972a, mattheiss1972b} and furthermore 
the effects of the structural change and electron correlation on the band structure have not been determined. 
In this Letter, we report on the temperature-dependent change of the
electronic structure of STO single crystals. We observed that the
Fermi surface of STO consists of three degenerate ellipsoids (\ie\
\dxy, \dyz, and \dzx) above \TC. As the temperature decreases
below \TC, we found that the \dxy\ band has lower minimum energy by
25 meV than the doubly degenerate \dyz\ and \dzx\ bands, which is in
the opposite direction to the theoretical predictions. 
The conduction band minimum gradually shifts in energy as
the STO experiences the second-order structural phase transition. 
In addition, the energy dispersion curves
near $\mathrm{\Gamma}$ point display a  \textit{peak-dip-hump} lineshape, where
the hump states can be attributed to a polaron formation.

STO single crystals with an atomically flat (001) surface of
TiO$_{2}$ termination were
prepared by chemical etching followed
by thermal annealing \cite{chang2009}. To introduce negative carriers  into
the samples, we annealed them in ultrahigh vacuum (UHV) ($<$ 1
$\times$ 10$ ^{-9}$ Torr) at $\sim$1000$^\circ$C for 30 min, 
upon which oxygen vacancies are are introduced with electron donation \cite{takizawa2009}. 
We are then able to examine the conduction band minimum structure over an energy range of $\sim$ 50 meV.
The carrier concentration was
estimated to be $n\sim  10^{20}$ cm$^{-3}$ from the size of Fermi
surface area (see below). ARPES was conducted at the Electronic
Structure Factory end station at beam line 7 of the Advanced Light
Source. The temperature was varied between 150$^\circ$K and 20$^\circ$K to span
the phase transition at \TC. The photon energy and the total energy
resolution (photons+electrons) were set to 95 eV and 30 meV, respectively, 
for all data in the present report.

\figone

We checked the structural phase transition of the STO sample using low energy electron diffraction (LEED) measurements. Fig.\ 1 shows two LEED pictures of the STO crystal taken at 300$^\circ$K to 20$^\circ$K. Over this wide temperature range, the evident diffraction spots imply both sufficient surface quality and a high enough electrical conductivity of the STO crystal for electron diffraction and spectroscopies. At certain electron beam energies, \eg\ 235 eV, the two (1,2) or (2,1) spots become evident at 20$^\circ$K, as indicated by arrows in Fig.\ 1(b). This fully reversible rearrangement of the diffraction spot intensity is attributed to the regular in-plane rotations of octahedral TiO$_{6}$ units in the long range order at low temperature, \ie\ cubic-to-tetragonal phase transition \cite{wang1973, krainyukova2000}.

\figtwo

Fig.\ 2 shows the Fermi surface and band structure maps of the STO collected at 150$^\circ$K and 20$^\circ$K. In the cubic phase at 150$^\circ$K, we find that the Fermi surface has three degenerate bands centered at $\mathrm{\Gamma}$ points and elongated along each principal axis as shown in Fig.\ 2 (a). The magnified Fermi surfaces in Fig.\ 2(b,d) present clear shapes of a circle and ellipse concentrated at points \cite{aiura2002}. The photoemission matrix element causes the intensity of each band to vary strongly from zone to zone. While no bands are visible at $\mathrm{\Gamma}_{00}=(0,0)$, the elliptic bands are strongest at $\mathrm{\Gamma}_{01}=(0,2\pi)$ and $\mathrm{\Gamma}_{10}=(2\pi,0)$ while the  circular band is strongest at $\mathrm{\Gamma}_{11}=(2\pi,2\pi)$. The complete Fermi surface is constructed from the sum of signals in $\mathrm{\Gamma}_{10}$, $\mathrm{\Gamma}_{11}$, $\mathrm{\Gamma}_{01}$ as shown in the inset in Fig.\ 2(a). The central circular band presents the
cross-section of the out-of-plane-oriented \dxy\ ellipsoidal pocket, which oscillates as the photon energy changes (not shown here). The elliptic bands, with an aspect ratio of about 2.4, represent 
cuts through the in-plane-oriented \dyz\ and \dzx\ ellipsoidal pockets. The elongation of the ellipsoids is much stronger than the findings of the Shubnikov de Haas oscillations \cite{gregory1979, uwe1985}, but is in agreement with the theoretical band model \cite{mattheiss1972b}. A similar elongation of the ellipsoidal bands may also be expected for other cubic perobskite compounds, such as KNiF$_{3}$, KMoO$_{3}$, KTaO$_{3}$ \cite{mattheiss1972a}.

Fig.\ 2(c,e) show the energy dispersion curves along the
horizontal line in the middle of Fig.\ 2(b,d), respectively.
Fig.\ 2(c) shows the intense circular band with the dim elliptic
bands aside, while Fig.\ 2(e) only shows the intense elliptic band
with the same minimum energy, \ie\ 46 meV below the Fermi level.
By comparing the area of the threefold degenerate ellipsoids compared to the area of a \BZ\, 
the carrier density is estimated to be 0.0085 per Ti atoms,
\ie\ 1.4 $\times$ 10$ ^{20}$ cm$^{-3}$. The effective mass can be
estimated to be \mel/\m0 = 1.2 (light electrons) and \meh/\m0\ =
7.0 (heavy electrons), where \m0\ is the free electron mass, from
the energy distribution curves (red dashed lines) of the circular and elliptic bands in
Fig.\ 2(c,e), respectively. 

\figthree

As temperature decreases across the cubic-to-tetragonal phase
transition, Mattheiss predicted the lifting of the degeneracy between \dxy, \dyz, \dzx\ 
ellipsoids by a transfer of the charge from \dxy\ to \dyz, \dzx\ pockets, 
corresponding to a shrinking of the former and an expansion of the latter. 
However, we observed the temperature dependence of
the conduction bands in the opposite direction. As shown in Fig.\
2(f,i), the elliptic bands shrink significantly at 20$^\circ$K. On
the other hand, the circular band expands a little, noticeable from
the guidelines (red dashed) and a slight increase in binding energy 
of the band minimum. The energy-dispersion curves clearly
show the temperature induced changes, as shown in Fig.\ 2(h,j).
The elliptic bands rise by 14 meV, but the circular band lowers by 11 meV. 
The total change in energy, 25 meV, is the same magnitude as predicted (20.7 meV), but of opposite sign \cite{mattheiss1972b}.

A more detailed understanding of the many-body interactions is obtained 
by analyzing the energy dispersion curves at \kx\ = $-2\pi$ and $-1.85\pi$ marked
by solid and dashed arrows in Fig.\ 2(h). As
shown in Fig.\ 3, the energy distribution curves show clear
peak-dip-hump structures, which are  associated with many-body
correlations \cite{damascelli2003}. As temperature decreases, the
peak-dip-hump structure becomes sharper indicating a substantial change in quasiparticle lifetime. Comparing the spectra at 20$^\circ$K 
and 150$^\circ$K in Fig.\ 3(a), the peak clearly displaces to lower
energy at lower temperature. The evolution of the bands demonstrates a continuous 
change of occupation with temperature. The gradual shift of the peak
with temperature confirms as expected, since the angle of tetragonal distortion 
increases continuously with temperature \cite{wang1973}.

As seen in Fig.\ 2(c,h), we found additional significant
electronic ("hump") states in the energy range near 170 meV. 
The hump below the three degenerate bands becomes discernible and lowers its energy
during cooling. The similar energy lowering of both the \dxy\ band and the hump 
during cooling implies the interaction between those two states. 
Since the hump is not visible at the (0, $-2\pi$) or ($-2\pi$, 0) points, 
it is clearly not associated with the \dyz\ or \dzx\ bands.
Comparing the energy range and the temperature dependence
with a recent optical study \cite{vanmechelen2008}, we can attribute
the hump states to polaron formation due to the electron-phonon
interaction. Similar polaronic bands have been observed 
in one-dimensional materials \cite{perfetti2001} and manganites \cite{mannella2005}

While the temperature dependence clearly implicates the polaronic origin for the hump states, we cannot rule out an additional contribution from plasmon satellites. The energy values of plasmon, \ie\ 60 meV for $n\sim 10^{20}$ cm$^{-3}$ \cite{gervais1993}, is similar to the energy separation from the hump to the peak. Further experiments on the doping dependence could rule out the plasmon because it would increase the separation between the peak and hump \cite{vonallmen1992}.

The renormalized band structure and these many-body interactions have not been available until the present work. They are important because they are the basis of STO's interesting properties such as superconductivity and thermoelectricity. In particular, the superconductivity has been treated only in the three-fold degenerate band model, which is clearly not applicable. Furthermore, the nature of the satellite bands, either polaronic or plasmonic, should be understood better by future calculations, as it is critical to establish an electron-phonon or electron-electron pairing mechanisms.

The precise measure of electronic structure is critical to understand the promising thermoelectricity observed in doped STO heterostructures. The thermoelectricity is directly related to the effective mass, which has been measured by different experimental methods \cite{frederikse1966, gregory1979, uwe1985, ohta2007} and rarely compared with calculations \cite{marques2003, wunderlich2008}. Differences between our observed value of effective mass and the calculated values suggests the need for improved electronic structure calculations and further investigation of STO with different doping. In addition, the thermoelectric properties have been previously associated with strong correlation \cite{ando1999}. The proposed polaron effects (Fig.\ 3) suggests that the electron-phonon coupling might play an important role for the thermoelectricity in doped STO. 

While the conduction band structure of the $n$-doped STO were previously extracted using the photoemission spectroscopy \cite{aiura2002, takizawa2009, haruyama1996, ishida2008} as well as
the magneto-transport \cite{gregory1979, uwe1985}, the precise evaluation of the Fermi surface shape, effective mass, and electron-phonon interaction from ARPES band structure measurements
is unique to the present study. The present work is only possible because of the very high energy and momentum resolutions of the experiments that also cover several \BZ s of STO. Future application of ARPES should be valuable to understand other interesting phenomena in STO
\cite{kan2005, kim2009} and STO-based heterostructures \cite{kozuka2009, ohtomo2004}.

In summary, we identified the band structure of STO, its modification due to polaronic effects and the evolution of these interactions with temperature during the structural phase transitions of electron-doped STO crystals.

The Advanced Light Source is supported by the Director, Office of
Science, Office of Basic Energy Sciences, of the US Department of
Energy under Contract No. DE-AC03-76SF00098. Y. J. C. and K. H.
acknowledge the support by the Max Planck Society.

\bibliography{stoPRL1}

\end{document}